# Constraining the Correlation Distance in Quantum Measurements

*Jean Schneider*
*LUTh – Paris Observatory*
*Jean.Schneider@obspm.fr*

## Abstract

Standard Quantum Physics states that the outcome of measurements for some distant entangled subsystems are instantaneously statistically correlated, whatever their mutual distance. This correlation presents itself as if there were a correlation at a distance with infinite speed. It is expressed by the Bell Theorem. It has been experimentally verified over distances up to 18 km with a time resolution of a few picosecond, which can be translated into an apparent effective correlation speed larger than $10^7$c.

The purpose of the present White Paper is to discuss the scientific interest and the feasibility to extend the correlation distance up to the Earth-Moon distance, i.e. $2\ 10^4$ times larger than in present experiments.

We are thus led to propose to install on the Moon a polarimter and a high performance photon detector with a high temporal resolution.

Such an exploratory experiment would provide new tests of Quantum Physics and could perhaps discriminate between standard Quantum Physics and for instance the Bohmian theory.

**Introduction**

Quantum Measurement is, with the irreversibility of time, one of the most profound problems of Physics. It has direct connections with philosophy since it raises the question of the nature of Knowledge and illustrates the debate between essentialism (knowledge unveils the secrets of nature) and constructivism (there are only observables and « nature » is nothing but an abstract construction).

Quantum physics is presently the most robust Fundamental Theory. Contrary to General Relativity which is at the verge to be modified if Dark Energy is confirmed, it has remain unchanged since its creation in the late 20's, in spite of several attempts to modify it (addition of a dissipative term in the Schrödinger equation (Hameroff-Nanopoulos), Spontaneous Localization, Bohm-de Broglie-Vigier subquantum medium). But quantum measurement raises the most profound question in Physics. Its paradoxical nature has appeared with the Born principle that the outcome is not predictible. This has as consequence that the outcome of measurements on two subsystems of a system are correlated in a non classical way. Non classical means here that the hypothesis of local hidden variables leads to predictions in contradiction with standard Quantum Physics.

**The problem**

There is no internal logical contradiction in Quantum Physics and up to now no contradiction with experimental results. The problem of Quantum Physics raised by several physicists rather resides in the contradictions between Quantum Measurement Theory and everyday life intuition and some philosophical prejudices. One of the paradoxes is the description of a measurement coming from the Von Meuman state vector reduction axiom. If the measuring apparatus is described as a physical system represented by a state vector and the measurement by an interaction represented by an interaction hamiltonian, the final state of the apparatus is a superposition of states representing the measurement results, contrary to the fact that single measurements have single outcomes.

This situation has led to several alternative theories such as the Bohm-Vigier theory of non local hidden variable carried by a « sub-quantume » medium (Bohm-Vigier) and the « spontaneous collapse » theory (Ghirardi, Rimini and Weber 1986). In a vein completely different from a

Bohmian-like Theory, let us mention the possibility of a gravitationally induced collapse of the state vector (Penrose 1996). It touches the rôle of classical space in Quantum Theory, one of the most profond aspects of Quantum Theory, This unclear relation between classical space and Quantum Physics was already pointed by N. Bohr's claims that « Quantum Physics transcends [classical] space and time ». This would have probable consequences for a quantum treatment of gravity.

**Present experimental test of Quantum Meausrement Theory**

According the standard Quantum Physics postulates, there is no limit in the above mentionned correlation distance. The correlation at-a-distance has been experimentally proven up to 18 km (Salart et al 2008). The question one may raise is « does the validity of the correlation at a distance hold for any distance up tu infinity, or is there a limiting distance? » A limiting distance would constitute a new Fundamental Constant and would force to change these postulates.

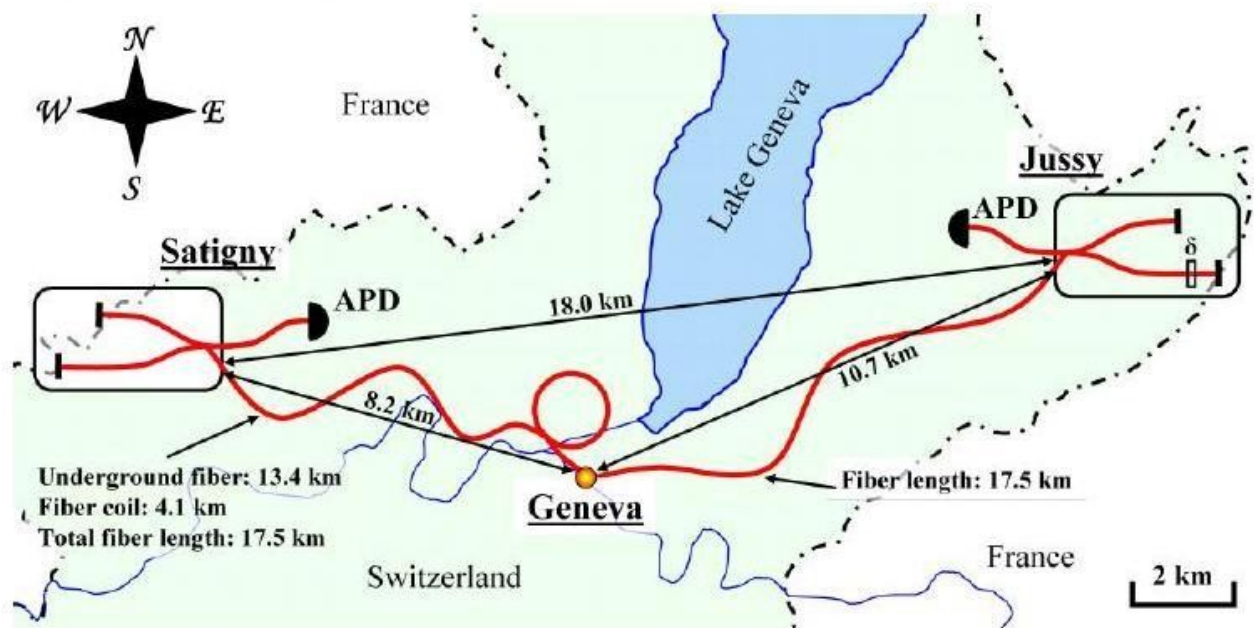

Illustration 1.: Salart et al (2008) Bell-type correlation experiment

It is not unreasonable to consider the possibility that at some distance instantaneous correlations may disappear. Are we indeed really ready to accept that it is valid up to infinity? In the Bohm-Vigier theory where correlations are carried by a « sub-quantum » medium, the only constraint is that the transportation speed is larger than the present experimental limit ($10^7$ c in case of the Salart et al (2008) experiment). There is no fundamental principle in the Bohm-Vgier theory requiring that this speed is infinite.

Salart et al (2008) express the correlation in terms of speed of some propagation, which would imply that « something » propagates. But from a conceptual point of view one has to dissociate a propagation speed V from a possible distance upper limit D. Four options are a priori possible:

- Correlation distance extended up to infinity, instanteneously (i.e. with zero time delay): it is the prediction of standard Quantum Theory, but also possible for instance in a Bohmian Theory.

- Instantaneous correlation distance limited to some value D (with for instance a correlation factor damping exponentially with a length scale D). The « speed of propagation » of correlation would then be infinite.

- Infinite correlation distance, but with a finite speed of propagation V.

- Finite correlation distance, with a finite speed of propagation V

A Bohmian-like Theory would be compatible with the last three options.

**What distance scale or « speed of propagation » of non standard correlations?**

One may wonder if there can be a priori theoretical estimates for a distance scale D or a speed of propagation V of non standard correlations. Playing only with the usual fundamental constants $h$, $c$ and $G$ can only give V = $c$ or V = $\infty$ and D = Planck length $10^{-33}$ cm. Playing in addition with less fundamental constants like the quark mass $m_q$ allows to multiply these values by any arbitrary power N of the dimensionless constant $Gm_q^2/hc \sim 10^{-39}$. For small values of N like -1 or +1 one gets V = $10^{+/-39}$ $c$ and D = $10^{-33+/-39}$ cm. Another source of a priori estimates for the distance scale D could come from possible non standard theories or phenomena like short-scale « fifth » force (D= a few cm to a few meter), the MOND theory as an alternative theory to dark matter (D ~ 1Mpc, Soussa and Woodward 2003), the Pioneer Anomaly (D ~ 10 AU) or the distance D = $c$T = 10 Mpc derived from the time scale T = $10^8$ yr for the « spontaneous collapse» (Ghirardi, Remini and Weber 1986).

As one can see, the span of possible a priori predictions has no solid constraint and only experiment may eventually provide a constraint.

**A proposed experimental approach.**

If, like in the three last above-mentionned options, the correlation does **not** extend to infinity with zero delay, the measured correlation must differ from the standard quantum prediction for a given time resolution $\Delta T$ beyond a certain distance D. This deviation from quantum predictions would be easier to detect (or vice-versa the confirmation of standard quantum prediction would gain in solidity) with increasing distance between the two photon detectors.

We therefore propose to extend as much as possible the distances at which quantum correlations are experimentally tested.

*On Earth,*

On Earth, beyond limitations given by the propagation of photons in optical fibers, one faces an absolution limitation, the finite Earth dimension. The largest distance between two points is 13,000 km. Here we propose to make experimental test of coorelations over larger distances made possible thanks to spacecrafts.

*Beyond Earth*

A few proposals have already proposed to test Bell inequalities in space, for instance by using the International Space Station (Ursin et al 2008, Kaltenbaek et al 2003, Pfennigbauer et al 2005).

Here we propose to use a **Lunar base** to make similar experiments. The gain in distance compared to the ISS would be a factor 1000.

   *An intermediaite step.*

One could just think of a set up similar to standard terrestrial set-ups, but with one arm pointing toward a mirror on the Moon, like for instance in the Lunar laser ranging experiments such as MéO at Observatoire de la Côte d'Azur. Both photon detection swould take place on Earth at a distance of at most of a few thousand kilometers. The detection itself would not take place on the Moon and according to standard Quantum Measurement theory, the correlation distance (i.e. the distance between the two detectors) would not be larger than a few thousand kilometers, in spite of a voyage of $2\times 350,000=700,000$ km for one of the two photons. Nevertheless, if in a Bohmian-like theory the correlation distance is shorter than a few hundred thousand km, the measured entanglement correlation would not agree with the prediction of standard Quantum Theory. In spite of not being a true space experiment, it would constitute a preliminary step.

*A final set-up*

In a final set-up, a powerful laser source is installed on Earth, with one branch pointing toward the Moon. In a future Lunar base a polarimeter and a photo detector for be installed on the Moon. The other branch, with a second polarimeter and detector would be on Earth. The experiment would consist in measuring the statistical correlation between photons counts as a function of the relative orientation of polarizers, like in the classical Salert et al (2008) experiment.

The ESA lunar lander with a possible launch in the 2017-20 timeframe (within the European Transportation and Human Exploration Preparatory Activities programme and the Global Exploration Strategy (GES)) would not be an appropriate frame since (in addition to the fact that the AO is closed) it would probably not permit a high technology fragile experiment.

Another more promising frame is the NASA Lunar Science Institute NLSI for which in Europe at least in France a NASA affiliated French Lunar Science Institute is in preparation.

***Beyond the Moon***

Beyond the Moon, Kaltenbaek et al (2003) have already proposed to make correlation experiments with a **Martian base**. The advantage is the gain in distance (a factor at least 200), But a complex Martian base required for such an experiment cannot be forseen before many decades.

Another possibility would perhaps be provided by the **LISA** project of the European Space Agency where laser links will exist between three spacecrafts separated by 5 millions km. Our proposal is then to use polarized lasers and to install orientable polarizers in two spacrafts and a source of correlated pairs of photons on the third spacecraft. Compared to Earth-Moon correlations the gain in distance would be a factor 20.

**Conclusion**

We invite FPRAT to recommend internal studies at ESTEC to investigate further the feasibility of the proposed experiments. If they were positive, they would open the greatest breakthrough in Physics since the invention of Quantum Theory. In particular it would help understanding the rôle of classical space in Quantum Physics, with possible consequences for the quantum treatment of gravity. Even if they were negative in the sense that they do not find a distance dependance of quantum correlations they would anyway improve space-based quantum communication.